\documentclass[aps,twocolumn]{revtex4}

\usepackage{graphicx}
\usepackage{bm}
\usepackage{epsfig}
\usepackage{amsmath}
\usepackage{amsthm}
\usepackage{amssymb} 

\usepackage{bbold}
\usepackage{rotating}

\begin{document}

\title{Collective Almost Synchronization in Complex Networks}

\author{M.  S.  Baptista$^1$, Hai-Peng Ren$^{2,1}$, J. C. M. Swarts$^{3}$, R. Carareto$^{4,1}$, H. Nijmeijer$^{3}$, C. Grebogi$^1$}
\affiliation{$^1$Institute for Complex
Systems and Mathematical Biology, University of
Aberdeen, SUPA, AB24 3UE Aberdeen, United Kingdom}
\affiliation{$^2$Department of Information and Control Engineering, Xi'an University of technology, 
5 Jinhua South Road, Xi'an, 710048, China}
\affiliation{$^3$Department of Mechanical Engineering, Dynamics and Control Group, 
Eindhoven University of Technology, WH 0.144,
P.O. Box 513, 5600 MB Eindhoven, The Netherlands}
\affiliation{$^4$Escola Politecnica,  
Universidade de S\~ao Paulo,
Avenida Prof. Luciano Gualberto, travessa 3, n. 158, 05508-900 S\~ao Paulo, SP, Brazil}

\begin{abstract}
  This work introduces the phenomenon of Collective Almost
  Synchronization (CAS), which describes a universal way of how
  patterns can appear in complex networks even for small coupling
  strengths.  The CAS phenomenon appears due to the existence of an
  approximately constant local mean field and is characterized by
  having nodes with trajectories evolving around periodic stable
  orbits. Common notion based on statistical knowledge would lead one
  to interpret the appearance of a local constant mean field as a
  consequence of the fact that the behavior of each node is not
  correlated to the behaviors of the others. Contrary to this common
  notion, we show that various well known weaker forms of
  synchronization (almost, time-lag, phase synchronization, and
  generalized synchronization) appear as a result of the onset of an
  almost constant local mean field.  
  If the memory is formed
  in a brain by minimising the coupling strength among neurons and
  maximising the number of possible patterns, then the CAS phenomenon
  is a plausible explanation for it. 
\end{abstract}
\maketitle

Spontaneous emergence of collective behavior is common in nature
\cite{bouchaud_MD2000,couzin_ASB2003, helbing_nature2000}. It is a
natural phenomenon characterized by a group of individuals that are
connected in a network by following a dynamical trajectory that is
different from the dynamics of their own.
Since the work of Kuramoto \cite{kuramoto_LNP1975}, the spontaneous
emergence of collective behavior in networks of phase oscillators with
full connected nodes or with nodes connected by some special
topologies \cite{acebron_RMP2005} is analytically well
understood.  Kuramoto considered a fully connected network of an
infinite number of phase oscillators. If $\theta_i$ is the variable
describing the phase of an oscillator $i$ in the network, and
$\overline{\theta}$ represents the mean field defined as
$\overline{\theta}=\frac{1}{N}\sum_{i=1}^N\theta_i$, collective
behavior appears in the network because every node becomes coupled to
the mean field.  Peculiar characteristics of this collective behavior
is that not only $\theta_i \neq \overline{\theta}$ but also nodes
evolve in a way that cannot be described by the evolution of only one
individual node, when isolated from the network.

In contrast to collective behavior, another widely studied behavior
of a network is when all nodes behave equally, and their evolution can
be described by an individual node when isolated from the network.
This state is known as complete synchronization
\cite{fujisaka_PTP1983}. If $x_i$ represents the state variables of an
arbitrary node $i$ of the network and $x_j$ of another node $j$, and
$\overline{x}$ represents the mean field of a network, complete
synchronization appears when $x_i=x_j=\overline{x}$, for all time. The
main mechanisms responsible for the onset of complete synchronization
in dynamical networks were clarified in
\cite{pecora_PRL1998,nijmeijer_PHYSICAD2009,nijmeijer_IEEE2011}. In
networks whose nodes are coupled by non-linear functions, such as
those that depend on time-delays \cite{nijmeijer_IEEE2011} or those
that describe how neurons chemically connect \cite{baptista_PRE2010},
the evolution of the synchronous nodes might be different from the
evolution of an individual node, when isolated from the network.
However, when complete synchronization is achieved in such networks,
$x_i=x_j=\overline{x}$.

In natural networks as biological, social, metabolic, neural networks, etc,
\cite{barabasi_RMP2002}, the number of nodes is often large but
finite; the network is not fully connected and heterogeneous. The
later means that each node has a different dynamical description or
the coupling strengths are not all equal for every pair of nodes, and
one will not find two nodes, say it $x_i$ and $x_j$, that have equal
trajectories. For such heterogeneous networks, as in
\cite{zhou_CHAOS2006,gardenes_chaos2011}, found in natural networks and in experiments
\cite{juergen_book}, one expects to find other weaker forms of
synchronous behavior, such as practical synchronization
\cite{femat_PLA1999}, phase synchronization \cite{juergen_book},
time-lag synchronization \cite{rosemblum_PRL1997}, and generalized
synchronization \cite{rulkov_PRE1995}.

We report a phenomenon that may appear in complex networks ``far
away'' from coupling strengths that typically produce complete
synchronization or these weaker forms of synchronization. However, the
reported phenomenon can be characterized by the same conditions used
to verify the existence of these weaker forms of synchronization.  We
call it Collective Almost Synchronization (CAS).  It is a consequence
of the appearance of an approximately constant local mean field and is
characterized by having nodes with trajectories evolving around stable
periodic orbits, denoted by $\mathbf{\Xi}_{p_i}(t)$, and regarded as a
CAS pattern. The appearance of an almost constant mean field is
associated with a regime of weak interaction (weak coupling strength)
in which nodes behave independently
\cite{jirsa_CN2008,batista_PRE2007}.  In such conditions, even weaker
forms of synchronization are ruled out to exist. But, contrary to
common notion based on basic statistical arguments, we show that
actually it is the existence of an approximately constant local mean
field that paves the way for weaker forms of synchronization (such as
almost, time-lag, phase, or generalized synchronization) to occur in
complex networks.

Denote all the $d$ variables of a node $i$ by ${\mathbf{x}}_i$, then
we define that this node presents CAS if the following inequality
\begin{equation}
|\mathbf{x}_i(t)-\mathbf{\Xi}_{p_i}(t-\tau_i)|  < \epsilon_i 
\label{CAS}
\end{equation}
is satisfied for {\it most of the time}. The double vertical bar $|\
|$ represents that we are taking the absolute difference between
vector components appearing inside the bars ($L1$ norm). $\epsilon_i$
is a small quantity, not arbitrarily small, but reasonably smaller
than the envelop of the oscillations of the variables
$\mathbf{x}_i(t)$.  $\mathbf{\Xi}_{p_i}(t)$ is the $d$-dimensional CAS
pattern. It is determined by the effective coupling strength $p_i$, a
quantity that measures the influence on the node $i$ of the nodes that
are connected to it, and the expected value of the local mean field at
the node $i$, denoted by $\mathbf{C}_i$. The local mean field, denoted
by $\overline{\mathbf{x}}_i$, is defined only by the nodes that are
connected to the node $i$.  The CAS pattern is the solution of a
simplified set of equations describing the network when
$\overline{\mathbf{x}}_i=\mathbf{C}_i$.  According to Eq.
(\ref{CAS}), if a node in the network presents the CAS pattern, its
trajectory stays intermittently close to the CAS pattern but with a
time-lag between the trajectories of the node and of the CAS pattern.
This property of the CAS phenomenon shares similarities with the way
complete synchronization appears in networks of nodes coupled under
time-delay functions \cite{nijmeijer_IEEE2011}. In such networks,
nodes become completely synchronous to a solution of the network that
is different from the solution of an isolated node of the network.
Additionally, the trajectory of the nodes present a time-lag to this
solution.

The CAS phenomenon inherits the three main characteristics of a
collective behavior: (a) the variables of a node $i$ ($\mathbf{x}_i$)
differ from both the mean field $\overline{\mathbf{x}}$ and the local
mean field $\overline{\mathbf{x}}_i$; (b) if the local mean fields of
a group of nodes and their effective coupling are either equal or
approximately equal, that causes all the nodes in this group to follow
the same or similar behaviors; (c) there can exist an infinitely large
number of different behaviors (CAS patterns).

If the CAS phenomenon is present in a network, other weaker forms of
synchronization can be detected. This link is fundamental when making
measurements to detect the CAS phenomenon.

In Ref.  \cite{femat_PLA1999}, the phenomenon of almost
synchronization is introduced, when a master and a slave in a
master-slave system of coupled oscillators have equal phases but
their amplitudes can be different.  If a node $i$ presents the CAS
phenomenon [satisfying Eq.  (\ref{CAS})] and $\tau_i=0$ in Eq.
(\ref{CAS}), then the node $i$ is almost synchronous to the pattern
$\mathbf{\Xi}_{p_i}$.

Time-lag synchronization \cite{rosemblum_PRL1997} is a phenomenon that
describes two identical signals, but whose variables have a
time-lag with respect to each other, i.e. $\mathbf{x}_i(t)=\mathbf{x}_j(t-\tau)$. In
practice, however, an equality between $\mathbf{x}_i(t)$ and $\mathbf{x}_j(t-\tau)$
should not be expected to be typically found, but rather
\begin{equation}
\mathbf{x}_i(t) \cong \mathbf{x}_j(t-\tau), 
\label{lag} 
\end{equation}
\noindent
meaning that there is not a constant $\tau$ that can be found such
that $\mathbf{x}_i(t) = \mathbf{x}_j(t-\tau)$.  Another suitable way
of writing Eq. (\ref{lag}) is by $|\mathbf{x}_i(t) -
\mathbf{x}_j(t-\tau)| \leq \gamma$. If two nodes $i$ and $j$ that
present the CAS phenomenon, have the same CAS pattern, and $\tau_i
\neq \tau_j \neq 0$, then
\begin{equation}
|\mathbf{x}_i(t) - \mathbf{x}_j(t-\tau_{ij})| \leq \epsilon_{ij}
\label{AS1}
\end{equation}
\noindent
or alternatively $\mathbf{x}_i(t) \cong \mathbf{x}_j(t-\tau_{ij})$,
for most of the time, $\tau_{ij}$ representing the time-lag between
$\mathbf{x}_i$ and $\mathbf{x}_j$. This means that almost time-lag
synchronization occurs for two nodes that present the CAS phenomenon
and that are almost locked to the same CAS pattern.  Even though nodes
that have equal or similar local mean field (which usually happens for
nodes that have equal or similar degrees) become synchronous with the
same CAS pattern (a stable periodic orbit), the value of their
trajectories at a given time might be different, since their
trajectories reach the neighborhood of their CAS patterns in different
places of the orbit.  As a consequence, we expect that two nodes that
exhibit the same CAS should present between themselves a time-lag
synchronous behavior.  For some small amounts of time, the difference
$|\mathbf{x}_i(t)-\mathbf{x}_j(t - \tau_{ij})|$ can be large, since
$\tau_{i} \neq \tau_j$ and $\epsilon_{i} \neq \epsilon_{j}$, in Eq.
(\ref{CAS}). The closer $\overline{\mathbf{x}}_i$ and
$\overline{\mathbf{x}}_j$ are to $\mathbf{C}_i$, the smaller is
$\epsilon_{ij}$ in Eq. (\ref{AS1}).

Phase synchronization \cite{juergen_book} is a phenomenon where the
phase difference, denoted by $\Delta \phi_{ij}$, between the phases of
two signals (or nodes in a network), $\phi_i(t)$ and $\phi_j(t)$,
remains bounded for all time
\begin{equation}
\Delta \phi_{ij}  = \left| \phi_i(i)-\frac{p}{q}\phi_j(t) \right| \leq S.
\label{PS}
\end{equation}
\noindent
In Ref. \cite{juergen_book} $S=2\pi$ and $p$ and $q$ are two rational
numbers.  If $p$ and $q$ are irrational numbers and $S$ is a
reasonably small constant, then phase synchronization can be referred
as to irrational phase synchronization \cite{baptista_PRE2004}. The
value of $S$ is calculated in order to encompass oscillatory systems
that possess either a time varying time-scale or a variable time-lag.
Simply make the constant $S$ to represent the growth of the phase in
the faster time scale during one period of the slower time scale.
Phase synchronization between two coupled chaotic oscillators was
explained as being the result of a state where the two oscillators
have all their unstable periodic orbits phase-locked
\cite{juergen_book}. Nodes that present the CAS phenomenon have
unstable periodic orbits that are locked to the stable periodic orbits
described by $\mathbf{\Xi}_i(t)$.  If $\mathbf{\Xi}_i(t)$ has a period
$P_i$ and the phase of this CAS pattern changes $D\phi_i$ within one
period, so the angular frequency is $\omega_i=D\phi_i/P_i$. If
$\mathbf{\Xi}_j(t)$ has a period $P_j$ and the phase of its CAS patter
changes $D\phi_j$ within one period, so the angular frequency is
$\omega_j=D\phi_j/P_j$. Then, the CAS patterns of these nodes are
phase synchronous by a ratio of $\frac{p}{q}=\omega_i/\omega_j$. Since
the trajectories of these nodes are locked to these patterns, the
nodes are phase synchronous by this same ratio, which can be rational
or irrational.  Assume additionally that, as one changes the coupling
strengths between the nodes, the expected value $\mathbf{C}_i$ of the
local mean field of a group of nodes remains the same. As a
consequence, as one changes the coupling strengths, both the CAS
pattern and the ratio $\frac{p}{q}=\frac{p_j D\phi_i}{p_i D \phi_j}$
remain unaltered, and the observed phase synchronization between nodes
in this group is stable under parameter alterations.


Consider a network of $N$ nodes with nodes connected diffusively (more
general networks are treated in the Supplementary Information)
described by
\begin{equation}
  \dot{\mathbf{x}}_i=\mathbf{F}_i(\mathbf{x}_i)+ \sigma \sum_{j=1}^N{\mathbf{A}_{ij}}{\mathbf{E}}(\mathbf{x}_j-\mathbf{x}_i), 
  \label{methods_network}
\end{equation}
\noindent
where $\mathbf{x}_i \in \Re^d$ is a d-dimensional vector describing the state
variables of the node $i$, $\mathbf{F}_i$ represents the dynamical system 
of the node $i$, and ${\mathbf{A}_{ij}}$ is the adjacent 
matrix. If $A_{ij}=1$, then, the node $j$ is connected
to the node $i$. ${\mathbf{E}}$ is the coupling function 
The degree of a node can be calculated by 
$k_i=\sum_{j=1}^NA_{ij}$. 

The CAS phenomenon  
appears when the local mean field of a node $i$, 
$\overline{\mathbf{x}}_i(t)=1/k_i \sum_{j}A_{ij}\mathbf{x}_j$,  is approximately constant
and $\overline{\mathbf{x}}_i(t) \approxeq \mathbf{C}_i$. Then, the equations for the
network can be described by
\begin{equation}
  \dot{\mathbf{x}}_i = \mathbf{F}_i(\mathbf{x}_i) - p_i E(\mathbf{x}_i) + p_i E(\mathbf{C}_i) + \mathbf{\delta}_i, 
\label{methods_network5}
\end{equation}
\noindent
where $p_i=\sigma k_i$ and the residual term is
$\mathbf{\delta}_i=p_i(\overline{\mathbf{x}}_i(t)-\mathbf{C}_i)$. The
CAS pattern of the node $i$ (a stable periodic orbit) is calculated in
the variables that produce a finite bounded local average field. If 
all components of $\mathbf{x}_i$ are bounded, then the CAS pattern is 
given by a solution of 
\begin{equation}
  \dot{\mathbf{\Xi}}_{p_i} = F_i(\mathbf{\Xi}_{p_i}) - p_i E(\mathbf{\Xi}_{p_i}) + p_i E(\mathbf{C}_i).  
\label{methods_network6}
\end{equation}
\noindent
which is just the same set of equations (\ref{methods_network5})
without the residual term. So, if $\overline{\mathbf{x}}_i(t) =
\mathbf{C}_i$, the residual term $\mathbf{\delta}_i=0$, and if Eq.
(\ref{methods_network6}) has no positive Lyapunov exponents
($\mathbf{\Xi}_{p_i}$ is a stable periodic orbit), then the node $x_i$
describes a stable periodic orbit.  If $\overline{\mathbf{x}}_i(t) -
\mathbf{C}_i$ is larger than zero but $\mathbf{\Xi}_{p_i}$ is a stable
periodic orbit, then the node $x_i$ describes a perturbed version of
$\mathbf{\Xi}_{p_i}$.  The closer $\overline{\mathbf{x}}_i$ is to
$\mathbf{C}_i$, the larger the time that Eq. (\ref{CAS}) is satisfied
at a given time. The more stable the periodic orbit is [the larger the
largest negative Lyapunov exponents of Eq.  (\ref{methods_network6})],
the longer Eq. (\ref{CAS}) is satisfied at a given time.

If the network has unbounded state variables (as it is the case of
Kuramoto networks \cite{kuramoto_LNP1975}), the CAS pattern is the
periodic orbit of period $T_i$ defined in the velocity space such that
$\dot{\mathbf{\Xi}}_{p_i}(t)=\dot{\mathbf{\Xi}}_{p_i}(t+T_i)$.

Notice that whereas Eqs. (\ref{methods_network}) and
(\ref{methods_network5}) represent a $Nd$-dimensional system, Eq.
(\ref{methods_network6}) has only dimension $d$.

The existence of this approximately constant local mean field is a
consequence of the Central Limit Theorem, applied to variables with
correlation (for more details, see Supplementary Information). The
expected value of the local mean field can be calculated by
\begin{equation}
  \mathbf{C}_i = _{\lim t \rightarrow \infty}\frac{1}{t}\int \overline{\mathbf{x}}_i(t) dt, 
\end{equation}
\noindent
where in practice we consider $t$ to be large, but finite. The larger
the degree of a node, the higher is the probability for the local mean
field to be close to an expected value and smaller its variance. If
the probability to find a certain value for the local mean field of
the node $i$ does not depend on the higher order moments of
$\overline{\mathbf{x}}_i(t)$, then this probability tends to be
Gaussian for sufficiently large $k_i$. As a consequence, the variance
$\mu^2$ of the local mean field is proportional to $k_i^{-1}$.

There are two criteria for the node $i$
to present the CAS phenomenon:
\begin{description}
\item[Criterion 1:] The Central Limit Theorem can be applied, i.e., 
  $\mu^2_i \propto k_i^{-1}$. Therefore, the larger the degree of a node, 
the smaller the variation of the local mean field $\overline{\mathbf{x}}_i(t)$ about 
its expected value $\mathbf{C}_i$.
\item[Criterion 2:] The CAS pattern $\mathbf{\Xi}_i(t)$ describes a
  stable periodic orbit.  The node trajectory can be considered to be
  a perturbed version of its CAS pattern.  The more stable the faster
  trajectories of nodes come to the neighborhood of the periodic
  orbits (CAS patterns), and the longer they stay around them.
\end{description}

Whenever the Central Limit Theorem applies, the random variables
involved are independent. But, the Central Limit Theorem can also be
applied to variables with correlation. If nodes that present the CAS
phenomenon are locked to the same CAS pattern, their trajectories
still arrive to the CAS pattern at different ``random'' times,
allowing for the Central Limit Theorem to be applied.  But the
time-lag between two nodes ($\tau_{ij}$) is approximately constant,
since the CAS pattern has a well defined period, and the trajectories
of these nodes are locked into it. The local mean field measured in a
node $i$ remains unaltered as one changes the coupling strength either
when the network has an infinite number of nodes (e.g. Kuramoto
networks) or the nodes have a symmetric natural measured (See Secs. C,
D, and E of Supplementary Information).  However, as we show in the
following example, the local mean field remains unaltered even when
the network has only a finite number of nodes and it has a natural
measure with no special symmetrical properties.

 
As an example to illustration how the CAS phenomenon appears in a
complex network, we consider a scaling-free network formed by, say,
$N=1000$ Hindmarsh-Rose neurons, with neurons coupled electrically.
The network is described by
\begin{eqnarray}
  \dot{x}_i &=&  y_i+3x_i^2-x_i^3-z_i+I + \sigma \sum_{j=1}^N A_{ij}(x_j -  x_i) \nonumber \\
  \dot{y}_i &=& 1-5x_i^2-y_i\label{HR} \\
  \dot{z}_i &=& -rz_i+4r(x_i+1.618)\nonumber 
\end{eqnarray}
where $I$=3.25 and $r$=0.005. The first coordinate of the equations
that describe the CAS pattern is given by
\begin{equation}
  \dot{\Xi}_{{x}_i} =  \Xi_{{y}_i} +3{\Xi}_{{x}_i}^2-{\Xi}_{{x}_i}^3-{\Xi}_{{z}_i}+I_i  - p_i{\Xi}_{{x}_i}  + p_i C_i. 
\label{methods_network6_HR}
\end{equation}

\begin{figure}[t]
\includegraphics[height=8.0cm,width=8.0cm]
{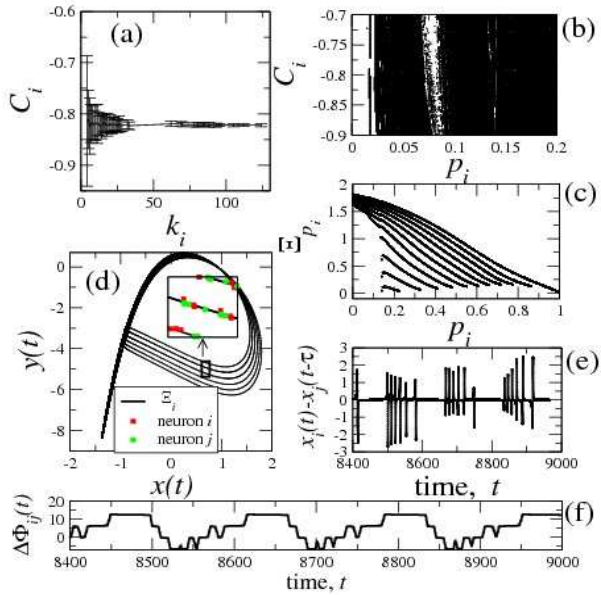}
\caption{{\footnotesize [Color online] (a) Expected value of the local mean field of
    the node $i$ against the node degree $k_i$. The error bar
    indicates the variance ($\mu^2_i$) of $\overline{x}_i$. (b) Black
    points indicate the value of $C_i$ and $p_i$ for Eq.
    (\ref{methods_network6_HR}) to present a stable periodic orbit (no
    positive Lyapunov exponents).  The maximal values of the periodic
    orbits obtained from Eq. (\ref{methods_network6_HR}) is shown in
    the bifurcation diagram in (c) considering $C_i=-0.82$ and
    $\sigma=0.001$.  (d) The CAS pattern for a neuron $i$ with degree
    $k_i$=25 (with $\sigma=0.001$ and $C=-0.82$). In the inset, the
    same CAS pattern of the neuron $i$ and some sampled points of the
    trajectory for the neuron $i$ and another neuron $j$ with degree
    $k_j=25$. (e) The difference between the first coordinates of the
    trajectories of neurons $i$ and $j$, with a time-lag of
    $\tau_{ij}=34.2$.  (f) Phase difference between the phases of the
    trajectories for neurons $i$ and $j$.}}
\label{figure2}
\end{figure}

\noindent
The others are given by $\dot{\Xi}_{{y}_i}=1-
\Xi_{{x}_i}^2-\Xi_{{y}_i}$,
$\dot{\Xi}_{{z}_i}=-r\Xi_{{z}_i}+4r(\Xi_{{x}_i}+1.618)$. In this
network, we have numerically verified that {\bf criterion 1} is
satisfied for neurons that have degrees $k \geq 10$ if $\sigma \leq
\sigma^*$, with $\sigma^* \cong 0.001$.  In Fig. \ref{figure2}(a), we
show the expected value $C_i$ of the local mean field of the first
coordinate ${x}_i$ of a neuron $i$ with respect to the neuron degree
(indicated in the horizontal axis), for $\sigma=0.001$. The error bar
indicates the variance of $C_i$ which fits to $\propto k_i^{-1.0071}$.
In (b), we show a parameter space to demonstrate that the CAS
phenomenon is a robust and stable phenomenon.  Numerical integration
of Eqs. (\ref{HR}) for $p_i \in [0.001,1]$ produces $C_i \in
[-0.9,0.7]$. We integrate Eq.  (\ref{methods_network6_HR}) by using
$C_i \in [-0.9,0.7]$ and $p_i \in [0,0.2]$, to show that the CAS
pattern is stable for most of the values. So, variations in $C_i$ of a
network caused by changes in a parameter do not modify the stability
of the CAS pattern calculated by Eq.  (\ref{methods_network6_HR}). For
$\sigma=0.001$, Eqs.  (\ref{HR}) yields many nodes for which
$\overline{x}_i \cong -0.82$. So, to calculate the CAS pattern for
these nodes, we use $C_i=-0.82$ and $\sigma=0.001$ in Eqs.
(\ref{methods_network6_HR}). The CAS pattern obtained, as we vary
$p_i$, is shown in the bifurcation diagram in (c), by plotting the
local maximal points of the CAS patterns.  {\bf Criterion 2} is
satisfied for most of the range of values of $p_i$ that produces a
stable periodic CAS pattern. A neuron that has a degree $k_i$ is
locked to the CAS pattern calculated by integrating Eqs.
(\ref{methods_network6_HR}) using $k_i\sigma=p_i$ and the measured
expected value for the local mean field, $C_i$.  In (d), we show the
periodic orbit corresponding to a CAS pattern associated to a neuron
$i$ with degree $k_i=25$ (for $\sigma$=0.001) and in the inset the
sampled points of the trajectories of this same neuron $i$ and of
another neuron $j$ that has not only equal degree ($k_j$=25), but it
feels also a local mean field of $C_j \cong -0.82$.  In (e), we show
that these two neurons have a typical time-lag synchronous behavior.
In (f), we observe $p/q=1$ phase synchronization between these two
neurons for a long time, considering that the phase difference remains
bounded by $S=6 \times 2\pi$ as defined in Eq.  (\ref{PS}), where the
number 6 is the number of spikings within one period of the slower
time-scale. In order to verify Eq.  (\ref{PS}) for all time, we need
to choose a ratio that is approximately equal to 1 ($p/q \cong 1$),
but not exactly 1 to account for slight differences in the local mean
field of these two neurons. Since $C_i$ depends on $\sigma$ for
networks that have neurons possessing a finite degree, we do not
expect to observe a stable phase synchronization in this network.
Small changes in $\sigma$ may cause small changes in the ratio $p/q$.
Notice however that Eq. (\ref{PS}) might be satisfied for a very long
time, for $p/q=1$. If neurons are locked to different CAS patterns
(and therefore have different local mean field), Eqs. (\ref{CAS}) and
(\ref{PS}) are both satisfied, but phase synchronization will not be
1:1, but with a ratio of $p/q$ (see Sec. E in Supplementary
Information for an example).

If neurons in this scaling-free network become completely synchronous,
it is necessary that $\sigma(N) \geq 2\sigma^{CS}(N=2)/|\lambda_2|$
(Ref.  \cite{pecora_PRL1998}). $\sigma^{CS}(N=2) \cong 0.5$
represents the value of the coupling strength when two bidirectionally
coupled neurons become completely synchronous. $\lambda_2=-2.06$
is the largest non-positive eigenvalue of the Laplacian matrix defined
as $A_{ij} - \mbox{diag}{(k_i)}$. So, $\sigma^{CS}(N) \geq 1/2.06
\cong 0.5$. The CAS phenomenon appears when $\sigma^{CAS}(N=1000)
\leq 0.001$, a coupling strength 500 times smaller than the one which
produces complete synchronization.  Similar conclusions would be
obtained when one considers networks of different sizes, with nodes
having the same dynamical descriptions and same connecting topology.


Concluding, in this work we introduce the phenomenon of Collective
Almost Synchronization (CAS), a phenomenon that is characterized by
having nodes possessing approximately constant local mean fields. The
appearance of an approximately constant mean field is a consequence of
a regime of weak interaction between the nodes responsible to place
the node trajectory around stable periodic orbits.
A network has the CAS phenomenon if the Central Limit Theorem can be
applied, and it exists an approximately constant mean field. In other
words, the CAS is invariant to changes in the value of the expected
value of the local mean field, that might appear due to parameter
alterations (e.g. coupling strength). If the expected value of the
local field changes, but the Central limit Theorem can still be
applied, nodes of the network will present the CAS phenomenon and the
observed weak forms of synchronization among the nodes might (or not)
be preserved.  As examples of how common this phenomenon could be, we
have asserted its appearance in a large networks of chaotic maps (see
supplementary information), Hindmarsh-Rose neurons, and Kuramoto
oscillators (see supplementary information).  In the Supplementary
Information, we also discuss that the CAS phenomenon is a possible
source of coherent motion in systems that are models for the
appearance of collective motion in social, economical, and animal
behaviour.

\section{Supplementary Information}

\subsection{CAS and generalized synchronization}

Generalized synchronization \cite{rulkov_PRE1995,abarbanel_PRE1996} is
a common behavior in complex networks
\cite{hung_PRE2008,guan_chaos2009,hu_chaos2010}, and should
be expected to be found typically.  This
phenomenon is defined as $x_i=\Phi(y_i)$, where $\Phi$ is considered
to be a continuous function. As explained in Refs.
\cite{rulkov_PRE1995,abarbanel_PRE1996}, generalized synchronization
appears due to the existence of a low-dimensional synchronous
manifold, often a very complicated and unknown manifold.

Recent works
\cite{zhou_CHAOS2006,ballerini_PNAS2008,pereira_PRE2010,gardenes_chaos2011}
have reported that nodes in the network that are highly connected
become synchronous. As shown in ref.  \cite{guan_chaos2009}, that is a
manifestation of generalized synchronization
\cite{rulkov_PRE1995,abarbanel_PRE1996} in complex networks. For a
fixed coupling strength among the nodes with heterogeneous degree
distributions and for the usual diffusively coupling configuration one
should expect that the set of hub nodes (highly connected nodes)
provides a skeleton about which synchronization is developed.  Reference
\cite{hramov_PRE2005} demonstrates how ubiquitous generalized
synchronization is in complex networks. It is shown that a necessary
condition for its appearance in
oscillators coupled in a driven-response (master-slave) configuration
is that the modified dynamics of the response system presents a stable
periodic behavior.  The modified dynamics is a set of equations
constructed by considering only the variables of the response system. In
a complex network, a modified dynamics of a node is just a system of
equations that contains only variables of that node.

An important contribution to understand why generalized
synchronization is a ubiquitous property in complex network is given
by the numerical work of Ref. \cite{guan_chaos2009} and the
theoretical work of Ref. \cite{hu_chaos2010}.  In Refs.
\cite{guan_chaos2009,hu_chaos2010} the ideas of Ref.
\cite{hramov_PRE2005} are extended to complex networks.  In
particular, the work of Ref. \cite{hu_chaos2010} shows that
generalized synchronization occurs whenever there is at least one node
whose modified dynamics is periodic.  All the nodes that have a stable
and periodic modified dynamics become synchronous in the generalized
sense with the nodes that have a chaotic modified dynamics.  The
general theorem presented in Ref.  \cite{hu_chaos2010} is a powerful
tool for the understanding of weak forms of synchronization or
desynchronous behaviors in complex networks.  However, identifying
the occurrence of generalized synchronization does not give much
information about the behavior of the network, since the function
that relates the trajectory among the nodes that are generalized
synchronous is usually unknown.  The CAS phenomenon allows one to
calculate, at least in an approximate sense, the equations of motion
that describes the pattern to which the nodes are locked to.  More
specifically, we can derive the set of equations governing, in an
approximate sense, the time evolution of the nodes, not covered by the
theorem in Ref. \cite{hu_chaos2010}.

Finally, if there is a node whose modified dynamics describes a stable
periodic behavior and its CAS pattern is also a stable periodic stable
behavior, then the CAS phenomenon appears when the network presents
generalized synchronization.

\subsection{CAS and other synchronous and weak-synchronous
  phenomena}



Consider a network of $N$ nodes described by
\begin{equation}
  \dot{\mathbf{x}}_i=\mathbf{F}_i(\mathbf{x}_i)+ \sigma \sum_{j=1}^N{\mathbf{A}_{ij}}{\mathbf{E}}[\mathcal{H}(\mathbf{x}_j-\mathbf{x}_i)] +\mathbf{\zeta}_i(t), 
  \label{network_sup}
\end{equation}
\noindent
where $\mathbf{x}_i \in \Re^d$ is a d-dimensional vector describing
the state variables of the node $i$, $\mathbf{F}_i$ is a
$d$-dimensional vector function representing the dynamical system 
of the node $i$, ${\mathbf{A}_{ij}}$ is the adjacent connection
matrix, ${\mathbf{E}}$ is the coupling function as defined in
\cite{pecora_PRL1998}, $\mathcal{H}$ is an arbitrary differentiable
transformation, and $\mathbf{\zeta}_i(t)$ is an arbitrary random
fluctuation.  Assume in the following that $\mathbf{\zeta}_i(t)=0$.

Assume that the nodes in the network (\ref{network_sup}) have equal
dynamical descriptions, i.e., $\mathbf{F}_i=\mathbf{F}$, that the network is fully
connected, so every node has a degree $k_i=N-1$, and that
$\mathcal{H}(\mathbf{x}_j-\mathbf{x}_i)=(\mathbf{x}_j-\mathbf{x}_i)$. We can rewrite it in terms of the
average field $\overline{\mathbf{x}}(t)=\frac{1}{N}\sum_{i=1}^N \mathbf{x}_i(t)$:
\begin{equation}
  \dot{\mathbf{x}}_i = \mathbf{F}_i(\mathbf{x}_i) - p_i \mathbf{E}(\mathbf{x}_i - \overline{\mathbf{x}}),   
  \label{network2_sup}
\end{equation}
where $p_i=\sigma k_i$. Therefore every node becomes ``decoupled''
from the network in the sense that their interaction is all mediated
by the average field.  Collective behavior is dictated by the
behavior of the average field and the individual dynamics of the
node. The linear stability of the network (\ref{network2_sup}) was
used in Ref.  \cite{zhou_CHAOS2006} as an approximation to justify how
desynchronous behavior about the average field can appear in complex
networks.  Notice that this assumption can only be rigorously
fulfilled if the network is fully connected and, therefore, it is
natural to understand why the desynchronous phenomena reported in Ref.
\cite{zhou_CHAOS2006} happens for nodes that are highly connected. One
can interpret the desynchronous behavior observed in Ref.
\cite{zhou_CHAOS2006} as an almost synchronization between a
node and the mean field $\overline{\mathbf{x}}$.

The differences between complete synchronization and synchronization
in the collective sense can be explained through the following
example. An interesting solution of Eq.  (\ref{network2_sup}) can be
obtained when $\overline{\mathbf{x}}=\mathbf{x}_i(t)$,
$\mathbf{x}_i(t)$ varying in time. In this case, the average field is
along the synchronization manifold. The network being completely
synchronous, all nodes having equal trajectories, and
$\mathbf{F}_i(\mathbf{x}_i(t))=\mathbf{x}_i(t)$. For such a special
network, collective behavior and complete synchronization are the
same. On the other hand, collective behavior typically appears when
the coupling term $\sigma E(\mathbf{x}_i - \overline{\mathbf{x}})$ is different from 
zero for most of the time and $\mathbf{F}_i(\mathbf{x}_i) \neq \mathbf{x}_i$, but there is a
majority of nodes with similar behavior. In this sense, the
desynchronous behaviors reported in Ref.  \cite{zhou_CHAOS2006} can
be considered as a collective phenomena that happens to parameters
close to the ones that yields complete synchronization.

To understanding when the CAS phenomenon occurs, consider the solution
of Eq. (\ref{network2_sup}) in the thermodynamics limit $N\rightarrow
\infty$ when $\overline{\mathbf{x}}$ is a constant in time, $\overline{\mathbf{x}}=C$.
For such a situation, the evolution of a node can be described by the
same following d-dimensional system of ODEs
\begin{equation}
  \dot{\mathbf{x}} = \mathbf{F}(\mathbf{x}) - p \mathbf{E}(\mathbf{x} - \mathbf{C}),  
  \label{network3_sup}
\end{equation}
\noindent
where $p=\sigma (N-1)$. If complete synchronization takes place, then
$\mathbf{F}_i(\mathbf{C})=0$, meaning that there can only exist
complete synchronization if all the nodes lock into the same stable
steady state equilibrium point, likely to happen if
$\mathbf{F}_i$ is the same for all the nodes.

Another possible network configuration that leads to
$\overline{\mathbf{x}}=\mathbf{C}$ happens when each node is only
weakly coupled (``independent'') with the others such that the Central
Limit Theorem could be applied. If the network has only a finite
number of nodes and $\overline{\mathbf{x}}(t)$ is not exactly constant
in time, but $\overline{\mathbf{x}}(t) \approxeq \mathbf{C}$, the
nodes still behave in the same predictable way if the dynamics
described by $\dot{\mathbf{x}} = \mathbf{F}(\mathbf{x}) - p
\mathbf{E}(\mathbf{x}) + p\mathbf{E}(\mathbf{C})$ is a sufficiently stable periodic
orbit.  This is how the CAS phenomenon appears in fully connected
networks.  All nodes become locked to the stable periodic orbit
described by $\dot{\mathbf{x}} = \mathbf{F}(\mathbf{x}) - p
\mathbf{E}(\mathbf{x}) + p \mathbf{E}(\mathbf{C})$.

Now, we break the symmetry of the network, allowing the nodes to be
connected arbitrarily to their neighbors. We still consider diffusive
linear couplings,
$\mathcal{H}(\mathbf{x}_j-\mathbf{x}_i)=(\mathbf{x}_j-\mathbf{x}_i)$.
The equations of such a network can be written as
\begin{equation}
  \dot{\mathbf{x}}_i = \mathbf{F}_i(\mathbf{x}_i) - p_i \mathbf{E}(\mathbf{x}_i) + p_i \mathbf{E}(\overline{\mathbf{x}}_i(t)), 
\label{network4_sup}
\end{equation}
\noindent
where $k_i$ is the degree of node $i$ with $k_l \leq k_m$, if $l<m$,
and $\overline{\mathbf{x}}_i(t)$ is the local mean field defined as 
\begin{equation}
\overline{\mathbf{x}}_i(t)=\frac{1}{k_i} \sum_{j=1}^N A_{ij} \mathbf{x}_j(t).
\label{local_mean_field_sup}
\end{equation}

Our main assumption is that the local mean field of a variable that is
bounded, either $\overline{\mathbf{x}}_i(t)$ or $\overline{\dot{\mathbf{x}}}_i(t)$,
exhibits small oscillations about an expected constant value $\mathbf{C}$.  In
other words, one can define a time average $\mathbf{C}$ by either
\begin{equation}
\mathbf{C}_i = \frac{1}{t} \int_0^t \overline{\mathbf{x}}_i(t) dt, 
\label{c_sup}
\end{equation} 
or 
\begin{equation}
\mathbf{C}_i = \frac{1}{t} \int_0^t \overline{\dot{\mathbf{x}}}_i(t)dt.
\label{c1_sup}
\end{equation} 
\noindent
Notice that ${\mathbf{x}}_i \in \Re^d$ (or $\dot{\mathbf{x}}_i \in
\Re^d$), and so does $\mathbf{C} \in \Re^d$. The CAS phenomenon
appears for a node that has at least one component of the local mean
field ($\overline{\mathbf{x}}_i$ or $\overline{\dot{\mathbf{x}}}_i$)
that is approximately constant. The appearance of this almost constant
value is a consequence of the Central Limit Theorem. For networks
whose nodes are described by only bounded variables, when calculating
the local mean field we only take into consideration the component
receiving the couplings from other nodes.  For networks of Kuramoto
oscillators that have one variable (the phase $\theta$) that is not
bounded, a constant local mean field appears in the component that
describes the instantaneous frequency ($\dot{\theta}_i$).

In Ref. \cite{hu_chaos2010}, it was shown that for chaotic networks
described by a system of equations similar to Eq. (\ref{network4_sup}),
generalized synchronization can appear if the modified dynamics
described by $\dot{\mathbf{x}}_i = \mathbf{F}_i(\mathbf{x}_i) - \sigma k_i \mathbf{E}(\mathbf{x}_i)$ of a
certain number of nodes are either stable equilibrium points
($\dot{\mathbf{x}}_i$=0) or they describe stable periodic solutions (limit
cycle). Generalized synchronization appears between the nodes that
have modified dynamics describing stable periodic states and the nodes
that have modified dynamics describing chaotic states.

To understand the phenomenon of {\it collective almost
  synchronization} (CAS), introduced in this work, consider that
$\mathcal{H}(\mathbf{x}_j-\mathbf{x}_i)=(\mathbf{x}_j-\mathbf{x}_i)$.
It is a phenomena that appears necessarily when
$\overline{\mathbf{x}}_i \approxeq \mathbf{C}_i$ or
$\overline{\dot{\mathbf{x}}}_i \approxeq \mathbf{C}_i$.  The equations
for the network can then be described by
\begin{equation}
  \dot{\mathbf{x}}_i = F_i(\mathbf{x}_i) - p_i \mathbf{E}(\mathbf{x}_i) + p_i \mathbf{E}(\mathbf{C}_i)+\mathbf{\delta}_i, 
\label{network5_sup}
\end{equation}
\noindent
where the residual term is 
$\delta_i=p_i(\overline{\mathbf{x}}_i - \mathbf{C}_i)$. This term is small most of the time
but large for some intervals of time; $\mathbf{\delta}_i(t)>0$ for all
time, but $\mathbf{\delta}_i(t) < \epsilon$ for most of the times.  Another
requirement for the CAS phenomenon to appear is that the CAS pattern
$\mathbf{\Xi}_i(t)$ of a node $i$ that is described by Eq.  (\ref{network5_sup})
ignoring the residual term
\begin{equation}
  \dot{\mathbf{\Xi}}_i = \mathbf{F}_i(\mathbf{\Xi}_i) - p_i \mathbf{E}(\mathbf{\Xi}_i) + p_i \mathbf{E}(\mathbf{C}_i).  
\label{network6_sup}
\end{equation}
\noindent
must be a stable periodic orbit. We define that a node presents collective almost synchronization (CAS)
if
\begin{equation}
|\mathbf{x}_i(t)-\mathbf{\Xi}_i(t-\tau_i)|  < \epsilon_i, 
\label{CAS_supplementary}
\end{equation}
{\it for most of the time},

Notice from Eq.  (\ref{network6_sup}) that for $p_i>0$, the CAS
pattern will not be described by $\mathbf{F}(\mathbf{x}_i)$ and
therefore does not belong to the synchronization manifold. On the
other hand, $\mathbf{\Xi}_i$ is induced by the local mean field as
typically happens in synchronous phenomenon due to collective
behavior. This property of the CAS phenomenon shares similarities
with the way complete synchronization appears in networks of nodes
coupled under time-delay functions \cite{nijmeijer_IEEE2011}. In such
networks, nodes become completely synchronous to a solution of the
network that is different from the solution of an isolated node of the
network.  Additionally, the trajectory of the nodes present a time-lag
to this solution.

To understand the reason why the CAS phenomenon appears when
$\mathbf{\Xi}_i(t)$ is a sufficiently stable periodic orbit, we study
the variational equation of the CAS pattern (\ref{network6_sup})
\begin{equation}
 \dot{\mathbf{\xi}}_i = [D\mathbf{F}_i(\mathbf{\xi}_i) - p_i\mathbf{E}]\mathbf{\xi}_i.     
\label{variational5}
\end{equation}
\noindent
obtained by linearizing Eq.  (\ref{network6_sup}) around
$\mathbf{\Xi}_i$ by making
$\mathbf{\xi}_i=\mathbf{x}_i-\mathbf{\Xi}_i$. This equation produces
no positive Lyapunov exponents. As a consequence, neglecting the
existence of the time-lag between $\mathbf{x}_i(t)$ and
$\mathbf{\Xi}(t)_i$, the trajectory of the node $i$ oscillates about 
$\mathbf{\Xi}_i$, and $\mathbf{x}_i-\mathbf{\Xi}_i \leq \epsilon_i$,
for most of the time, satisfying Eq.  (\ref{CAS_supplementary}), where
$\epsilon_i$ depends on $\mathbf{\delta}_i$.  If there are two nodes
$i$ and $j$,  which feel similar local mean fields, $\mathbf{\Xi}_i
\approxeq \mathbf{\Xi}_j$, then $\mathbf{x}_i \approxeq \mathbf{x}_j$,
for most of the time.

To understand why the nodes that present CAS have also between them a
time-lag type of synchronization, integrate Eq.
(\ref{network5_sup}), using Eq. (\ref{network6_sup}), to obtain
\begin{equation}
\mathbf{x}_i(t) = \int_{0}^t [\dot{\mathbf{\Xi}}_i(t) + \mathbf{\delta}_i(t)]dt.  
\label{distance1}
\end{equation}
\noindent
This integral is not trivial in the general case. But we have a simple
phenomenological explanation for its solution. When the CAS pattern is
sufficiently stable, the asymptotic time limit state of the variable
$\mathbf{x}_i(t)$ is the CAS pattern $\mathbf{\Xi}_i(t)$. But due to
the residual term $\mathbf{\delta}_i(t)$, the trajectory of
$\mathbf{x}_i(t)$ arrives in the neighborhood of $\mathbf{\Xi}(t)$ at
time $t$ with a time-lag. As a result, nodes that are collectively
almost synchronous obey Eq.  (\ref{CAS_supplementary}).  In addition,
two nodes that present CAS have also a time-lag between their
trajectories for the same reason.  There is an extra contribution to
the time-lag between the trajectories of two nodes if their initial
conditions differ.

Phase synchronization \cite{juergen_book} is a phenomena where the
phase difference, denoted by $\Delta \phi_{ij}$ between the phases of
two signals (or nodes in a network), $\phi_i(t)$ and $\phi_j(t)$,
remains bounded for all time
\begin{equation}
\Delta \phi_{ij}  = \left| \phi_i(i)-\frac{p}{q}\phi_j(t) \right| \leq S, 
\label{PS_sup}
\end{equation}
\noindent
where $S=2\pi$, and $p$ and $q$ are two rational numbers
\cite{juergen_book}.  For coupled chaotic oscillators one can also
find irrational phase synchronization \cite{baptista_PRE2004}, where
Eq. (\ref{PS_sup}) can be satisfied {\it for all time} with $p$ and
$q$ irrational. $S$ is a reasonably small constant, that can
be larger than 2$\pi$ in order to encompass oscillatory systems that
either have a time varying time-scale or whose time-lag varies
in time.  This bound can be simply calculated by making $S$ to
represent the growth of the phase in the faster time scale after one
period of the slower time scale.

The link between the CAS phenomenon and phase synchronization 
can be explained by thinking that it is a synchronous phenomenon
among the nodes that is mediated by their CAS patterns.  The phase
of the periodic orbit of the CAS pattern of the node $i$ grows as
$\tilde{\phi}_i(t)=\omega_i t + \xi_i(t) + \phi_i^0$ and of the node
$j$ grows as $\tilde{\phi}_j(t)=\omega_j t + \xi_j(t) + \phi_j^0$. The
quantities $\phi_i^0$ and $\phi_j^0$ are displacements of the phase
caused by the existence of time-lag, and $\xi_i(t)$ and $\xi_j(t)$ are
small fluctuations. For $t \rightarrow \infty$ these can be neglected
and we have that
\begin{equation}
\frac{\tilde{\phi}_i(t)}{\tilde{\phi}_j(t)}=\frac{\omega_i}{\omega_j}=\frac{p}{q}, 
\label{lock_frequencies}
\end{equation}
\noindent
where $\omega_i=\lim_{t \rightarrow
  \infty}\frac{\tilde{\phi}_i(t)}{t}$ gives the average frequency of
oscillation of the CAS pattern of node $i$, and $p$ and $q$ are
two real numbers.

The phase of the nodes can be written as a function of the phase of
the periodic orbits of the CAS pattern. So, $\phi(t)_i=
\tilde{\phi(t)}_i + \delta \phi_i(t)$ and $\phi(t)_j=
\tilde{\phi(t)}_j + \delta \phi_j(t)$, $\delta_i(t)$ represents a
variation of the phase of the node $i$ with respect to the phase of
the CAS pattern, and depends on the way the phase is defined
\cite{pereira_PLA2007}. The phase difference $\Delta \phi_{ij}(t)$,  as
written in Eq. (\ref{PS_sup}), becomes equal to
$|t(q\omega_i-p\omega_j)+q\delta_i(t)-p\delta\phi_j(t)|$. But, from
Eq. (\ref{lock_frequencies}), $q\omega_i-p\omega_j=0$, and therefore,
$\Delta \phi_{ij}(t) \leq \max{(q\delta\phi_i(t)-p\delta\phi_j(t))}$. But
since the node orbit is locked to the CAS pattern, $\Delta
\phi_{ij}(t)$ is always a small quantity. 

In practice, for networks composed by a finite number of nodes, we do
not expect that the quantities $\delta \phi_i(t)$ and $\delta
\phi_j(t)$ to remain small for all the time. The reason is that the
CAS pattern can only be approximately calculated and in general we do
not know the precise real value of the local mean field. However, our
simulations show that these quantities remain small for time intervals
that comprise many periods of oscillations of the node trajectories.
For networks having an expected value of the mean field $\mathbf{C}_i$
that is independent on the coupling strength $\sigma$, the ratio $p/q$
does not change as one changes the value of $\sigma$, and then phase
synchronization is stable under a parameter variation. For the network
of Kuramoto oscillators, Eq. (\ref{PS_sup}) can be verified for all
time with a value of $p/q$ that remains invariant as one changes
$\sigma$.

Assume for now that the nodes have equal dynamics, so
$\mathbf{F}_i=\mathbf{F}$. If a node $i$ with degree $k_i$ has a
periodic CAS pattern that is sufficiently stable under Eq.
(\ref{variational5}), all the nodes with degrees close to $k_i$ 
also have similar CAS patterns that are sufficiently stable under Eq.
(\ref{variational5}). Node $i$ is locked to $\mathbf{\Xi}_i$ and node
$j$ is locked to $\mathbf{\Xi}_j$.  But since $\mathbf{\Xi}_i$ is
approximately equal to $\mathbf{\Xi}_j$, thus, $\mathbf{x}_i
\cong \mathbf{x}_j$, for most of the time.  So, if the pattern
solution is sufficiently stable, the external noise
$\mathbf{\zeta}_i(t)$ can be different from zero, and still have
similar trajectories for that interval of time. The same argument
remains valid if $\mathbf{F}_i \neq \mathbf{F}_j$, as long as the CAS
pattern is sufficiently stable.

In Ref.  \cite{pereira_PRE2010}, synchronization was defined in terms
of the node $\mathbf{x}_N$ that has the largest number of connections,
when $\mathbf{x}_i(t) \cong \mathbf{x}_N$ (which is equivalent to
stating that $|\mathbf{x}_i(t) - \mathbf{x}_N| < \epsilon$), where
$\mathbf{x}_N$ is assumed to be very close to the synchronization
manifold $\mathbf{s}$ defined by $\dot{\mathbf{s}}=\mathbf{F}(\mathbf{s})$.
This type of synchronous behavior was shown to exist in scaling free
networks whose nodes have equal dynamics and that are linearly
connected. This was called hub synchronization.

The link between the CAS phenomenon with the hub synchronization
phenomenon \cite{pereira_PRE2010}, and generalized synchronization can
be explained as in the following. It is not required for nodes that
present the CAS phenomenon for their error dynamics $\mathbf{x}_j -
\mathbf{x}_i$ to be small. But for the following comparison, assume
that $\mathbf{\vartheta}_{ij} = \mathbf{x}_j - \mathbf{x}_i$ is small
so that we can linearise Eq.  (\ref{network4_sup}) about another node
$j$.  Assume also that $\mathbf{F}_i=\mathbf{F}$.  The variational
equations of the error dynamics between two nodes $i$ and $j$ that
have equal degrees are described by
\begin{equation}
  \dot{\mathbf{\vartheta}}_{ij} = [D\mathbf{F}(\mathbf{x}_i) - p_iE]\mathbf{\vartheta}_{ij} + 
  \mathbf{\eta}_i. 
\label{variational6}
\end{equation}
\noindent
In Ref. \cite{pereira_PRE2010}, hub synchronization exists if Eq.
(\ref{variational6}), neglecting the coupling term $\mathbf{\eta}_i$,
has no positive Lyapunov exponents. That is another way of stating
that hub synchronization between $i$ and $j$ occurs when the
variational equations of the modified dynamics
$[\dot{\mathbf{x}}_i=\mathbf{F}(\mathbf{x}_i)-p_i E(\mathbf{x}_i)]$
presents no positive Lyapunov exponent. In other words, in order to
have hub synchronization it is necessary that the modified dynamics of
both nodes be describable by stable periodic oscillations. Hub
synchronization is the result of a weak form of generalized
synchronization, defined in terms of the linear stability of the error
dynamics between two highly connected nodes.  Unlike generalized
synchronization, hub synchronization offers a way to predict, in an
approximate sense, the trajectory of the synchronous nodes.

In contrast, the CAS phenomenon appears when the CAS pattern, which is
different from the solution of the modified dynamics, becomes
periodic. Another difference between the CAS and the hub
synchronization phenomenon is that whereas $\overline{\mathbf{x}}_i
\approxeq \mathbf{C}$ in the CAS phenomenon, $\overline{\mathbf{x}}_i
\approxeq {\mathbf{x}}_i$ in the hub synchronization, in order for
$\mathbf{\eta}_i$ to be very small, and $\mathbf{x}_i$ to be close to
the synchronization manifold. So, whereas hub synchronization can be
interpreted as being a type of practical synchronization
\cite{femat_PLA1999}, CAS is a type of almost synchronization.

In the work of Refs. \cite{politi_PRE2006,politi_PRL2010}, it was
numerically reported a new desynchronous phenomenon in complex
networks. The network has no positive Lyapunov exponents but it
presents a desynchronous non-trivial collective behavior. A possible
situation for the phenomenon to appear is when $\mathbf{\delta}_i$ and
$\mathbf{C}_i$ in Eq. (\ref{network5_sup}) are either zero or sufficiently small
such that the stability of the network is completely determined by Eq.
(\ref{variational5}), and this equation produces no positive Lyapunov
exponent. Assume now that $p_i$ in Eq. (\ref{network6_sup}) is
appropriately adjusted such that the CAS pattern for every node $i$ is
a stable periodic orbit.  The variational Eqs. (\ref{variational5})
for all nodes have no positive Lyapunov exponents.  If additionally,
$\overline{\mathbf{x}}_i(t) \approxeq \mathbf{C}$, then the network in Eq.
(\ref{network4_sup}) possesses no positive Lyapunov exponent.
Therefore, networks that present the CAS phenomenon for all nodes
might present the desynchronous phenomenon reported in Refs.
\cite{politi_PRE2006,politi_PRL2010}. The CAS phenomenon becomes
different from the phenomenon of
Refs.\cite{politi_PRE2006,politi_PRL2010} if for at least one node,
Eq. (\ref{network6_sup}) produces a chaotic orbit.

To understand the occurrence of CAS in networks formed by heterogeneous
nodes connected by nonlinear functions such as networks of Kuramoto
oscillators, we rewrite the Kuramoto's network model in terms of the
local mean field, $\overline{\theta}_i=\frac{1}{k_i}
\sum_{j=1}^NA_{ij} \theta_j$. Using the coordinate transformation
\begin{equation}
  \frac{1}{k_i}\sum_{j=1}^NA_{ij}\exp{^{\mathbb{j}(\theta_j-\theta_i)}}=
  \tilde{r}_i \exp{^{\mathbb{j}(\overline{\theta}_i-\theta_i)}}, 
\end{equation}
\noindent
the  dynamics of the node $i$ is described by 
\begin{equation}
  \dot{\theta_i} = \omega_i + p_i \tilde{r}_i sin(\overline{\theta}_i-\theta_i). 
  \label{kuramoto2_sup}
\end{equation}

The phase ${\theta_i}$ is not a bounded variable and therefore we
expect that typically $\overline{\theta}_i$ has not a well defined
average.  But, $\overline{\dot{\theta_i}}(t)$ is bounded and has a
well defined average value which is an approximately constant quantity
($C_i$) for nodes in networks with sufficiently large number of
connections and with sufficiently small coupling strengths.  When
$\overline{\dot{\theta}_i} \cong C_i$, the node $i$ has the propensity
to exhibit the CAS phenomenon, and the CAS pattern is calculated by
Eq.  (\ref{kuramoto2_sup}) considering that
$\overline{\theta_i}=C_it$.  Notice that
$\overline{\theta_i}=\overline{\dot{\theta}_i}t \cong C_it$.

Phase synchronization between two nodes in the networks of Eq.
(\ref{kuramoto2_sup}) is stable under parameter variations (coupling
strength in this case) if these nodes present the CAS phenomenon.
There is irrational (rational) phase synchronization if
$\frac{\overline{\dot{\theta_i}}}{\overline{\dot{\theta_j}}}$ is
irrational (rational). If nodes are sufficiently ``decoupled'' we
expect that
$\frac{\overline{\dot{\theta_i}}}{\overline{\dot{\theta_j}}} \approxeq
\omega_i/\omega_j$. Phase
synchronization will be rational whenever nodes with different natural
frequencies become locked to Arnold tongues's, induced by the coupling
$p_i\tilde{r}_isin(\overline{\theta}_i-\theta_i)$.

There is a special solution of Eq. (\ref{kuramoto2_sup}) that produces a
bounded state in the variable ${\theta_i}$ when the network is
complete synchronous to an equilibrium point.  In such case,
$\overline{\theta}_i$ becomes constant, and Eq.  (\ref{kuramoto2_sup}) has
one stable equilibrium $\theta_i = \arcsin{\left( \frac{\omega_i}{p_i}
  \right)}$, obtained when $p_i > \omega_i$. But, the local mean field
becomes constant due to complete synchronization and not due to the
fact that the nodes are ``decoupled''. These conditions do not produce
the CAS phenomenon.

We take the thermodynamics limit when the network has infinite nodes
with infinite degrees. $C_i$ calculated using Eq. (\ref{c1_sup}) does
not change as one change the coupling $\sigma$, since
$\overline{\dot{\theta}_i}= \lim_{k_i,N \rightarrow \infty}
\frac{1}{k_i}[\sum_{j=1}^N A_{ij}(\omega_i + p_i \tilde{r}_i
sin(\overline{\theta_i}-\theta_i))]$=$\lim_{k_i,N \rightarrow \infty}
\frac{1}{k_i}[\sum_{j=1}^N A_{ij}\omega_j] + [\sum_{j=1}^N
A_{ij}(\sigma \tilde{r}_j sin(\overline{\theta_j}-\theta_i))]$ =
$\frac{1}{k_i}[\sum_{j=1}^N A_{ij}\omega_j] + \sigma \sum_{j=1}^N
A_{ij}(\tilde{r}_j sin(\overline{\theta_j}-\theta_i))$.  But, if nodes
are sufficiently decoupled $\sum_{j=1}^N A_{ij}(\tilde{r}_j
sin(\overline{\theta_j}-\theta_i))$ approaches zero, and therefore,
$C_j$ only depends on the natural frequencies:
$\overline{\dot{\theta}_i} = C_i = \frac{1}{k_i}[\sum_{j=1}^N
A_{ij}\omega_j]$.

Assume that there are two nodes, $i$ and $j$, and that for most of the
time $\Xi_i \approxeq \Xi_j$. Then, for most of the time it is also
true that $\Xi_i-\theta_i \approxeq \Xi_j - \theta_j$, which allow us
to write that $sin(\Psi_j-\theta_j)-sin(\Psi_i-\theta_i) \approxeq
\cos{(\Psi_i-\theta_i)}[(\Psi_j-\theta_j)- (\Psi_j-\theta_j)]
\approxeq \cos{(\Psi_i-\theta_i)}[\theta_j -\theta_j]$. Since $\Psi_i
\approxeq \theta_i$, then $\cos{(\Psi_i-\theta_i)} \approxeq 1$ and
$sin(\Psi_j-\theta_j)-sin(\Psi_i-\theta_i) \approxeq [\theta_j
-\theta_j]$. Defining the error dynamics between the two nodes to be
$\xi_{ij}=\theta_j-\theta_i$, we arrive that
\begin{equation}
\dot{\xi}_{ij} \approxeq (\omega_j-\omega_i) - p_i \xi_{ij}.
\end{equation}
\noindent
Therefore, it implies that we expect to find two nodes having the same
similar CAS behavior when both the local mean field is close and when
the difference between their natural frequencies $(\omega_j-\omega_i)$
is small.

The CAS phenomenon can also appear in a system of driven particles
\cite{vicsek_PRL1995} that is a simple but powerful model for the
onset of pattern formation in population dynamics
\cite{couzin_ASB2003}, economical systems \cite{gregoire_physicaD2003}
and social systems \cite{helbing_nature2000}. In the work of Ref.
\cite{vicsek_PRL1995}, it was assumed that individual particles were
moving at a constant speed but with an orientation that depends on the
local mean field of the orientation of the individual particles within
a local neighborhood and under the effect of additional external
noise.  Writing an equivalent time-continuous description of the
Vicsek particle model \cite{vicsek_PRL1995}, the equations of motion
for the direction of movement of a particle $i$, can be written as
\begin{equation}
\dot{\mathbf{x}}_i = - \mathbf{x}_i + \overline{\mathbf{x}}_i + \Delta \mathbf{\theta}_i, 
\label{continuous_vicsek}
\end{equation}
\noindent
where $\overline{\mathbf{x}}_i$ represents the local mean field of the
orientation of the particle $i$ within a local neighborhood and
$\Delta \mathbf{\theta}_i$ represents a small noise term. When $\overline{\mathbf{x}}_i$
is approximately constant, the CAS pattern is 
described by a solution of $\dot{\mathbf{x}}_i = - \mathbf{x}_i + \overline{\mathbf{x}}_i$,
which will be a stable equilibrium point as long as $\Delta \mathbf{\theta}_i$
is sufficiently small. From the Central Limit Theorem,
$\overline{\mathbf{x}}_i$ will be approximately constant as long as the
neighborhood considered is sufficiently large or the density of
particles is sufficiently large. 

\subsection{About the expected value of the local mean field: the Central Limit Theorem}

The Theorem states that, given a set of $t$ observations, each set of
observation containing $k$ measurements ($x_1,x_2,x_3,x_4,\ldots,
x_k$), the sum $S_N=\sum_{i=1}^k x_i(N)$ (for $N=1,2,\ldots,t$), with
the variables $x_i(N)$ drawn from an independent random process that
has a distribution with finite variance $\mu^2$ and mean
$\overline{x}$, converges to a Normal distribution for sufficiently
large $k$. As a consequence, the expected value of these $t$
observations is given by the mean $\overline{x}$ (additionally,
$\overline{x}=\frac{1}{t}\sum_{N=1}^t S_N$), and the variance of the
expected value is given by $\frac{\mu^2}{k}$.  The larger the number
$k$ of variables being summed, the larger is the probability with
which one has a sum close to the expected value. There are many
situations when one can apply this theorem for variables with some
sort of correlation \cite{hilhorst_BJP2009}, as it is the case for
variables generated by deterministic chaotic systems with strong
mixing properties, for which the decay of correlation is exponentially
fast. In other words, a deterministic trajectory that is strongly
chaotic behaves as an independent random variable in the long-term.
For that reason, the Central Limit Theorem holds for the time average
value $\overline{x}(t)$ produced by summing up chaotic trajectories
from nodes belonging to a network that has nodes weakly connected.
Consequently, the distribution of
$\overline{x}_i(t)=\frac{1}{N}\sum_jA_{ij}x_j(t)$ for node $i$ should
converge to a Gaussian distribution centered at
$C_i=\frac{1}{t}\int_0^t\overline{x}_i(t)dt$ as the degree of the node
is sufficiently large.  In addition, the variance $\mu^2_i$ of the
local mean field $\overline{x}(t)_i$ decreases proportional to
$k_i^{-1}$, as we have numerically verified for networks of
Hindmarsh-Rose neurons ($\mu^2_i \propto k_i^{-1.0071}$) and
networks of Kuramoto oscillators ($\mu^2_i \propto k_i^{-1.055}$).

If the network has no positive Lyapunov exponents, we still expect to
find an approximately constant local mean field at a node $i$, as long
as the nodes are weakly connected and its degree is sufficiently
large. To understand why, imagine that every node in the network stays
close to a CAS pattern and one of its coordinates is described by
$sin(\omega_it)$. Without loss of generality we can make that every
node has the same frequency $\omega_i=\omega$.  The time-lag property
in the node trajectories, when they exhibit the CAS pattern, results
in that every node is close to $sin(\omega_i t)$ but they will have a
random time-lag in relation to the CAS pattern (due to the
decorrelated property between the node trajectories). So, the selected 
coordinate can be described by $sin(\omega t + \phi^0_i)+\delta_i(t)$,
where $\phi^0_i$ is a random initial phase and $\delta_i(t)$ is a
small random term describing the distance between the node trajectory
and the CAS pattern. Neglecting the term $\delta_i(t)$, the
distribution of the sum $\sum_{i=1}^k sin(\omega t +
\phi^0_i)$ converges to a normal distribution with a variance
that depends on the variance of $sin(\phi^0_i)$.

From previous considerations, if the degree of some of the nodes tend
to infinite, the variance of the local mean field for those nodes
tends to zero and, in this limit, the residual term $\delta_i$ in Eq.
(\ref{network5_sup}) is zero and the local mean field of these nodes
is a constant. As a consequence, the node is perfectly locked
with the CAS pattern ($\epsilon=0$ in Eq. (\ref{CAS_supplementary})).

\subsection{CAS in a network of coupled maps}

\begin{figure}[t]
\includegraphics[height=8.0cm,width=8.0cm]
{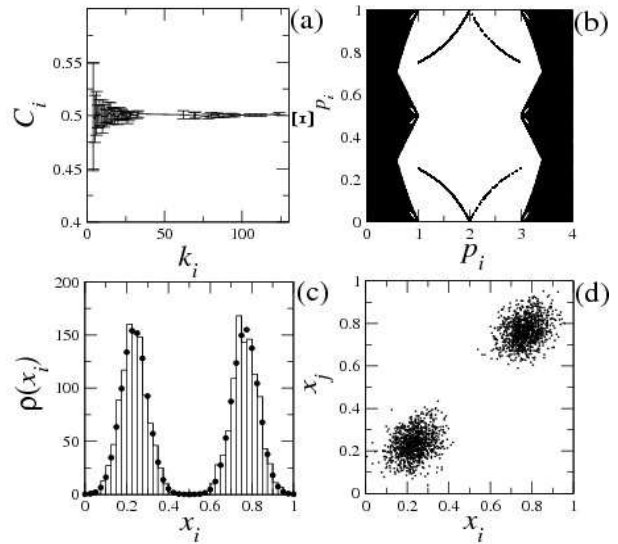}
\caption{{\footnotesize (a) Expected value of the local mean field of
    the node $i$ against the node degree $k_i$. The error bar
    indicates the variance ($\mu^2_i$) of $\overline{x}_i$. (b) A
    bifurcation diagram of the CAS pattern [Eq.
    (\ref{network6_sup})] considering $C_i=0.5$. (c) Probability
    density function of the trajectory of a node with degree $k_i$=80
    (therefore, $p_i=\sigma k_i=1.3$, $\sigma=1.3/80$). (d) A return
    plot considering two nodes ($i$ and $j$) with the same degree
    $k_i=k_j=$80.}}
\label{figure1}
\end{figure}

As another  example to illustrate how the CAS phenomenon appears in a
complex network, we consider a network of maps whose node dynamics is
described by $F_i(x_i)=2x_i$ mod(1). The network composed, say, by
$N=1000$ maps, is represented by $x_{i}^{(n+1)} =
F_i(x_i^{(n)})+\sigma \sum_{j=1}^N A_{ij}(x_j^{(n)}-x_i^{(n)})$
mod(1), where the upper index $n$ represents the discrete iteration
time, and $A_{ij}$ is the adjacency matrix of a scaling-free network.
The map has a constant probability density.  When such a map is
connected in a network, the density is no longer constant, but still
symmetric and having an average value of 0.5. As a consequence, nodes
that have a sufficient amount of connections ($k \geq 10$) feel a
local mean field, say, within $[0.475,0.525]$, (deviating of 5$\%$
about $C_i$=0.5) and $\mu^2_i \propto k_i^{-1}$ ({\bf criterion 1}),
as shown in Fig.  \ref{figure1}(a). Therefore, such nodes have
propensity to present the CAS phenomenon.  In (b) we show a
bifurcation diagram of the CAS pattern, $\Xi_i$, obtained from Eq.
(\ref{network6_sup}) by using $C_i=C=0.5$, as we vary $p_i$.
Nodes in this network that have propensity to present the CAS
phenomenon will present it if additionally $p_i \in [1,3]$; the CAS
pattern is described by a period-2 stable orbit ({\bf criterion 2}).
This interval can be calculated by solving $|2 - p_i| \leq 1$.  In (c)
we show the probability density function of the trajectory of a node
that present the CAS phenomenon. The density is centered at the
position of the period-2 orbit of the CAS pattern and for most of the
time Eq.  (\ref{CAS_supplementary}) is satisfied.  The filled circles are fittings
assuming that the probability density is given by a Gaussian
distribution.  Therefore, there is a high probability that
$\epsilon_i$ in Eq.  (\ref{CAS_supplementary}) is small. In (d) we show a plot of
the trajectories of two nodes that have the same degree which is equal
to 80.  We chose nodes which present no time-lag between their
trajectories and the trajectory of the pattern. If there was a
time-lag, the points in (d) would not be only aligned along the
diagonal (identity) line, but they would also appear off-diagonal.

\subsection{CAS in the Kuramoto network}

An illustration of this phenomenon in a network composed by nodes
having heterogeneous dynamical descriptions and a nonlinear coupling
function is presented in a random network of $N$=1000 Kuramoto
oscillators. We rewrite the
Kuramoto network model in terms of the local mean field,
$\overline{\theta}_i=\frac{1}{k_i} \sum_{j=1}^NA_{ij} \theta_j$. Using
the coordinate transformation 
$  \frac{1}{k_i}\sum_{j=1}^NA_{ij}\exp{^{\mathbb{j}(\theta_j-\theta_i)}}=
  \tilde{r}_i \exp{^{\mathbb{j}(\overline{\theta}_i-\theta_i)}} 
$, the  dynamics of node $i$ is described by 
\begin{equation}
  \dot{\theta_i} = \omega_i + p_i \tilde{r}_i sin(\overline{\theta}_i-\theta_i),  
  \label{kuramoto3}
\end{equation}
where $\omega_i$ is the natural frequency of the node $i$, taken from
a Gaussian distribution centered at zero and with standard deviation of
4. If $\tilde{r}_i$=1, all nodes coupled to node $i$ are
completely synchronous with it. If $\tilde{r}_i$=0, there is no
synchronization between the nodes that are coupled to the node
$i$. Since the phase is an unbounded variable, the CAS phenomenon
should be verified by the existence of an approximate constant local
mean field in the frequency variable $\dot{\theta_i}$.  If
$\overline{\dot{\theta}_i}(t) \cong C_i$, which means that
$\overline{\theta_i}=\overline{\dot{\theta}_i}t \cong C_it$, then Eq.
(\ref{kuramoto3}) describes a periodic orbit (the CAS pattern),
regardless the values of $\omega_i$, $p_i$, and $\tilde{r}_i$, since
it is an autonomous two-dimensional system; chaos cannot exist.
Therefore, {\bf criterion 2} is always satisfied in a network of
Kuramoto oscillators. We have numerically verified that {\bf
  criterion 1} is satisfied for this network for $\sigma \leq
\sigma^{CAS}(N=1000)$, where $\sigma^{CAS}(N=1000) \cong 0.075$.
Complete synchronization is achieved in this network for $\sigma \geq
\sigma^{CS} = 1.25$.  So, the CAS phenomenon is observed for a
coupling strength that is 15 times smaller than the one that
produces complete synchronization.

\begin{figure}[t]
\includegraphics[height=8.0cm,width=8.0cm]
{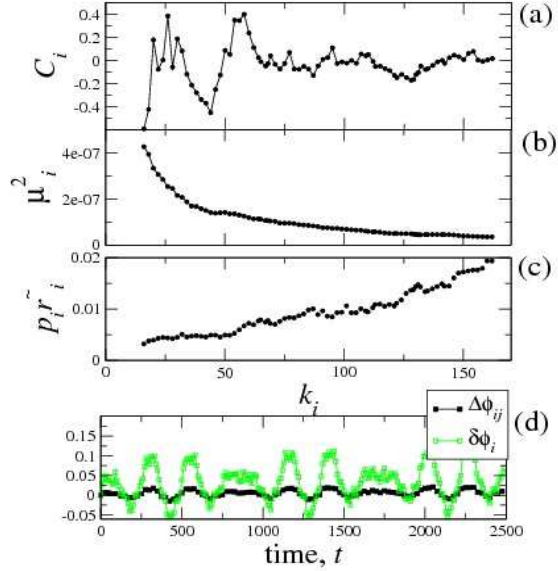}
\caption{{\footnotesize Results for $\sigma=0.001$. (a) Expected value
    of the local mean field $\overline{\dot{\theta}_i}$ of a node with
    degree $k_i$. (b) The variance $\mu^2_i$ of the local mean field.
    (c) Relationship between the value of $p_i \tilde{r}_i$ and
    $k_i$}. (d) Phase difference $\Delta
  \phi_{ij}=\theta_i-p/q\theta_j$ between two nodes, one with degree
  $k_i=96$ and the other with degree $k_j=56$; the phase difference
  $\delta \phi_i = \theta_i - \Xi_{\theta_i}$ between the phases of
  the trajectory of the node $i$ with degree $k_i=96$ and the phase of
  its CAS pattern.}
\label{figure3}
\end{figure}

For the following results, we choose $\sigma=0.001$.  Since the
natural frequencies have a distribution centered at zero, it is
expected that, for nodes with higher degrees, the local mean field
is close to zero (see Fig.  \ref{figure3}(a)).  In (b), we show the
variance of the local mean field of the nodes with degree $k_i$. The
fitting produces $\mu^2_i \propto k_i^{-1.055}$ ({\bf criterion 1}).
In (c), we show the relationship between the value of $p_i
\tilde{r}_i$ and the value of the degree $k_i$. In order to
calculate the CAS pattern of a node with degree $k_i$, we need to use
the value of $p_i \tilde{r}_i$ (which is obtained from this figure)
and the measured $C_i$ as an input in Eq.  (\ref{kuramoto3}). We pick
two arbitrary nodes, $i$ and $j$, with degrees $k_i=96$ and $k_j=56$,
respectively, with natural frequencies $\omega_i \approxeq -5.0547$
and $\omega_j \approxeq -5.2080$. In (d), we show that phase
synchronization is verified between these two nodes, with $p/q =
\omega_i/\omega_j$. We also show the phase difference $\delta \phi_j = \theta_j
- \Xi_{\theta_j}$ between the phases of the trajectory of the node $i$
with degree $k_j=96$ and the phase of its CAS pattern, for a time
interval corresponding to approximately 2500/$P$ cycles, where the
period of the cycles in node $i$ is calculated by
$P=\frac{2\pi}{5.0547}$. Phase synchronization between nodes $i$ and
$j$ is a consequence of the fact that the phase difference between the
nodes and their CAS patterns is bounded.

In the thermodynamic limit, when a fully connected network has an
infinite number of nodes, $C_i$ does not change as one changes the
coupling $\sigma$, since it only depends on the mean field of the
frequency variable ($\overline{\dot{\theta}}$).  As a consequence, if
there is the CAS phenomenon and phase synchronization between two
nodes with a ratio of $p/q$ for a given value of $\sigma$, changing
$\sigma$ does not change the ratio $p/q$. Therefore phase
synchronization is stable under alterations in $\sigma$. Phase
synchronization will be rational and stable whenever nodes with
different natural frequencies $\omega_i$ become locked to Arnold
tongues \cite{jensen,arnold_tongue} induced by the coupling
$p_i\tilde{r}_isin(\overline{\theta}_i-\theta_i)$.
    
There is a special solution of Eq. (\ref{kuramoto3}) that produces a
bounded state in the variable ${\theta_i}$ when the network is
complete synchronous to an equilibrium point.  In such case,
$\overline{\theta}_i$ becomes constant, and Eq.  (\ref{kuramoto3}) has
one stable equilibrium $\theta_i = \arcsin{\left( \frac{\omega_i}{p_i}
  \right)}$, obtained when $p_i > \omega_i$. But, the local mean field
becomes constant due to complete synchronisation and not due to the
fact that the nodes are ``decoupled''. These conditions do not produce
the CAS phenomenon.

\subsection{Preserving the CAS pattern in different networks: a way to
  predict the onset of the CAS phenomenon in larger networks}

Consider two networks, $n_1$ and $n_2$, whose nodes have equal
dynamical descriptions, the network $n_1$ with $N_1$ nodes and the
network $n_2$ with $N_2$ nodes ($N_2>N_1$), and two nodes, $i$ in the
network $n_1$ and $j$ in the network $n_2$. Furthermore, assume that
both nodes have stable periodic CAS patterns ({\bf criteria 1} is
satisfied), and assume that the nodes have sufficiently large degrees
such that the local mean field of node $i$ is approximately equal to
node $j$. Then the CAS pattern of node $i$ will be approximately the
same as the one of node $j$ if
\begin{equation}
  \sigma^{CAS}(n_1)k_i(n_1) = \sigma^{CAS}(n_2)k_j(n_2).
\label{predicting_CAS}
\end{equation} 
\noindent
$\sigma^{CAS}(n_1)$ and $\sigma^{CAS}(n_2)$ represent the
largest coupling strengths for which the variance of the local mean
field of a node decays with the inverse of the degree of the node
({\bf criterion 2} is satisfied) in the networks, respectively, and
$k_i(n_1)$ and $k_j(n_2)$ are the degrees of the nodes $i$ and $j$,
respectively. In other words, the CAS phenomenon occur in the
network if $\sigma \leq \sigma^{CAS}$.

Therefore, if $\sigma^{CAS}(N_1)$ is known, $\sigma^{CAS}(N_2)$ can be
calculated from Eq. (\ref{predicting_CAS}). In other words, if the CAS
phenomenon is observed at node $i$ for $\sigma \leq
\sigma^{CAS}(N_1)$, the CAS phenomenon will also be observed at node $j$
for $\sigma(n_2) \leq \sigma^{CAS}(n_2)$, where $\sigma^{CAS}(n_2)$
satisfies Eq. (\ref{predicting_CAS}).

\textbf{Acknowledgment} MSB acknowledges the partial financial support
of the Northern Research Partnership. HPR acknowledges the partial
financial support of NSFC Grant 60804040.


\end{document}